\documentclass[11pt]{article}
 \usepackage{epsf}
\def\thefootnote{\fnsymbol{footnote}}

\newcommand{\eq }{\begin{equation}}
\newcommand{\en}{\end{equation}}
\newcommand{\eqa}{\begin{eqnarray}}
\newcommand{\ena}{\end{eqnarray}}

\newcommand{\bra}{\langle}
\newcommand{\ket}{\rangle}

\newcommand{\ij}{\langle ij \rangle}
\newcommand{\bx}{{\bar\sqcup}}

\newcommand{\NP}[1]{Nucl.\ Phys.\ {\bf #1}}

\newcommand{\PR}[1]{Phys.\ Rev.\ {\bf #1}}
\newcommand{\PRL}[1]{Phys.\ Rev.\ Lett.\ {\bf #1}}

\begin{document}
\begin{titlepage}
\vskip0.5cm
\begin{flushright}
%{\tt draft}
DFTT 24/99\\
\end{flushright}
\vskip0.5cm
\begin{center}
{\Large\bf Thermal operators in Ising Percolation. }
\end{center}
\vskip 1.3cm
\centerline{
M. Caselle\footnote{e--mail: caselle@to.infn.it}
 and F. Gliozzi\footnote{e--mail:  gliozzi@to.infn.it}}
 \vskip 1.0cm
 \centerline{\sl   Dipartimento di Fisica
 Teorica dell'Universit\`a di Torino}
 \centerline{\sl Istituto Nazionale di Fisica Nucleare, Sezione di Torino}
 \centerline{\sl via P.Giuria 1, I-10125 Torino, Italy}
 \vskip .4 cm

\begin{abstract}
We discuss a new cluster representation for the internal energy and
the specific heat of the d-dimensional Ising model , obtained by
studying the percolation mapping of an Ising model with an arbitrary set
of antiferromagnetic links.
Such a representation relates the thermal operators  to the topological
properties of the Fortuin-Kasteleyn clusters of  Ising percolation and is a
powerful tool to get new exact relations on the topological structure of
FK clusters of the Ising model defined on an arbitrary graph.
\end{abstract}
\end{titlepage}

\setcounter{footnote}{0}
\renewcommand{\thefootnote}{\arabic{footnote}}
\section{Introduction}
\label{intro}
It is well known that the Ising model can be mapped into a percolation
problem~\cite{fk,ck}.
In its original formulation this mapping is based on the
 identification of the mean value of the magnetization of the Ising model with
 the size of a (suitably defined) percolating cluster. Such identification
 is highly
 non-trivial and  allows for a geometric characterization of
 several statistical observables near the critical point, a fact which has
 greatly improved our understanding of the Ising model and inspired new
 powerful algorithms~\cite{sw,wolf} to simulate the model.
  This mapping can be
 extended also to the thermal sector of the model and cluster
 representations of, for instance, the internal energy and the specific heat
 can be easily constructed. However these representations, in contrast to
 the case of magnetization, are rather trivial and do not add any new geometrical or
 topological information to the model.

In this paper we shall discuss a new cluster representation
for the thermal sector of the Ising model. To this end we shall need first to
 extend the percolation mapping to the case of the Ising
 model frustrated by some antiferromagnetic links~\cite {clmp} and
 then we use this mapping to extract a useful piece of information on the
 structure of the Fortuin Kasteleyn (FK) clusters of the unfrustrated model.

The main feature of our new representation  is
 that  it relates the thermal operators of the model (and in particular
 the internal energy and the specific heat)  to the topological properties
 of the clusters of a typical configuration of Ising percolation.
As a consequence one can write down exact relations on the
topological structure of the FK cluster.

The simplest of these new  relations deals with the concept of
non-cutting or {\sl black} bond. A bond of a FK cluster is said to be
black if its cancellation does not split the cluster in two disjoint
parts. We shall prove (see sect.\ref{thermal}) that in the Ising model
defined on an arbitrary
d-dimensional lattice with $N$ links at the coupling $\beta$, the mean
number $ \bra N_B\ket$ of black bonds is related to the internal energy
$E$ by
\eq
E=\bra N_B\ket \frac 1{\sinh 2\beta} \, +\,N\, \tanh\beta\, ~~.
\en
Combining this with the well known expression (see eq.(\ref{ie1})
below)
\eq
E=\bra N_G\ket\,\frac{{\rm e }^\beta}{\sinh\beta}-N~~,
\en
where $\bra N_G\ket$ is the mean number of bonds of FK graphs
associated the configurations of Ising percolation, yields
a simple, exact relation between $\bra N_B\ket$  and $ \bra N_G \ket$.
Similar relations can be found for the specific heat .

Actually these new relations are true for the Ising model defined on
any arbitrary graph, however in
extracting our results we shall always assume for sake of
simplicity that the model be defined on a regular lattice in $d=2,3,\dots$
dimensions with periodic boundary conditions.

In the process of constructing our new representation we shall
discuss two issues which are rather interesting in themselves.

First, we shall construct a scheme to classify the bonds
of a cluster on the basis of their topological properties (see
sect. \ref{links} below).
The standard classification \cite {perco}, which splits the bonds in three classes
(which are commonly denoted with the three colors  red,blue and green)
  refers to the bonds of the (infinite)
 percolating cluster. Our new representation of thermal operators suggests
 a slightly different bond partition, based on the connection properties
 of {\sl all} the clusters. Within  our classification scheme the
topological properties of the bonds are unambiguously identified by the class
to which they belong.

Second, we shall show in sect. \ref{duality}, using duality, that in the
 Ising model all the correlators  of an even
number of spins (and hence invariant under the $Z_2$ symmetry of the model)
can be expressed as ratios of partition functions with a suitable set
of antiferromagnetic links.

 This paper is organized as follows.
The first part of the paper is composed by an introductory section
on the Ising model and the percolation map (sect. \ref{ising}) which
will allow us to fix notations and to make the paper as self-contained as
possible. In   sect. \ref{links} we discuss
the topological properties of the clusters. In sect. \ref{frustration}
we deal with the Ising model in presence of
antiferromagnetic links and  duality
transformation and sect. \ref{fruma}  is devoted to
the extension of the percolation mapping to the frustrated Ising
model. In these last two
sections we have collected most of our new results which are then used in
sect. \ref{thermal} to construct new
 cluster representations for the internal energy and
specific heat. Finally sect. \ref{conclu} is devoted to some concluding remarks.

\section{Ising Model}
  \label{ising}
The   $d$ dimensional  Ising  model
is defined by the Hamiltonian:
\eq
 H(J,h') = - J\sum_{\bra n,m\ket} s_n s_m ~+~ h' \sum_n s_n\; ,
\label{Sspin}
\en
where
 the field variable $s_n$ takes the values $-1$ and $+1$;
 $n$ labels the sites of the lattice (denoted with $\Lambda$ in the following)
 which we assume to be a $d$ dimensional
 simple (hyper)cubic lattice of size $L$ with periodic boundary conditions,
 however a large part of our considerations are valid for a Ising
 model define on an arbitrary graph.
The notation
$\bra n,m\ket$  indicates that the sum is taken on  nearest neighbour sites
only.
The partition function is defined as usual by
\eq
Z=\sum_{s_n=\pm1}~e^{-\beta~H(J,h')}
\en
where  $\beta\equiv \frac{1}{kT}$.
Plugging eq.(\ref{Sspin}) in the definition of $Z$ and assuming the usual
conventions: $J=1$ and $h=\beta h'$ we obtain:
\eq
Z(\beta,h)=\sum_{s_n=\pm1}~e^{\beta~\sum_{\bra n,m\ket} s_n s_m ~+~ h \sum_n s_n}~~.
\label{part}
\en
For $h=0$ and $d\geq 2$ the phase diagram of the
 model is composed by two phases separated by a second order phase transition
 located at
 $\beta_c\equiv \frac{1}{kT_c}$. In the high temperature phase the $Z_2$
 symmetry of the model is preserved and the magnetization is zero, in the low
 temperature phase the $Z_2$
 symmetry is spontaneously broken and the magnetization becomes different from
 zero.

\subsection{Mapping to a percolation model}
\label{map}
In this section we shall discuss the mapping in the case $h=0$. The
extension to a non zero magnetic field is straightforward and can be
found for instance in~\cite{hc}.

The Ising partition function of eq.(\ref{part}) with
$h=0$ can be rewritten as:
\eq
Z(\beta)=e^{N\beta}\sum_{s_i=\pm1}\prod_{\bra ij\ket }
\left[e^{-2\beta}+(1-e^{-2\beta})\delta(s_i,s_j)\right]
\label{part2}
\en
 where $N\equiv dL^d$ is the number of links in the
 lattice and the $\delta$
function takes the value $\delta(s_i,s_j)=1$ when the
two arguments coincide and zero otherwise.
Expanding the products in eq.(\ref{part2}) one finds
\eq
Z(\beta)=e^{N\beta}\sum_{G}\sum_{s_i=\pm1}
\left[\prod_{\bra ij\ket \in G}
p\,\delta(s_i,s_j)\right](1-p)^{N-N(G)}
\label{part3}
\en
where $p=1-e^{-2\beta}$, $G$ denotes an arbitrary subgraph
of the lattice and $N(G)$ is the number of links of $G$.
In general $G$ will be composed by
several connected components ( FK clusters in the following).
Let us call $C(G)$
the number of clusters in the graph $G$.
 Notice that among the clusters one has to consider
also those with one site only.

Summing on the spin configurations in eq.(\ref{part3}) we end up with
\eq
Z(\beta)=e^{N\beta}\sum_{G}~p^{N(G)}(1-p)^{N-N(G)}~2^{C(G)}
\label{part4}
\en
which can be interpreted as the partition function of a percolation model
with bond probability $p$ and with a weight $2$ for each independent cluster.
In this framework the magnetization transition of the Ising model becomes a
percolation transition, located at the percolation threshold $p_c\equiv
1-e^{-2\beta_c}$.

For $p>p_c$ an infinite, percolating cluster exists.  The density of sites belonging to
 this percolating cluster, which is zero below $p_c$,
can be used as order parameter for the percolation transition. It can be shown
that it exactly coincides with the magnetization density
 of the original Ising model.

\vskip .2 cm
The internal energy in the percolation framework can be constructed
 by taking the logarithmic derivative of eq.(\ref{part4}) with
respect to $\beta$. The result is
\eq
E= \frac{2\bra N(G)\ket }{p}-N~~~,
\label{ie1}
\en
where the mean value is taken with respect to the measure
of eq.(\ref{part4}).
It is interesting to compare this result with the standard definition of
internal energy:
\begin{equation}
E =\frac{\partial}{\partial \beta} log Z(\beta,h=0)=
\bra \sum_{\bra n,m\ket } s_n s_m \ket
\label{intene}
\end{equation}
If we denote with $N_+$ the number of links which join spins
with the same sign in a given
configuration and $N_-\equiv N-N_+$ the number of those which join spins
with opposite sign then eq.(\ref{intene}) can be rewritten as:
\eq
E= \bra N_+\ket  - \bra N_-\ket  = 2\bra N_+\ket  -N~~.
\label{ie2}
\en
By comparing eq.s (\ref{ie1}) and (\ref{ie2}) we see that
\eq
p\bra N_+\ket= \bra N(G)\ket~~~,
\label{ie3}
\en
which gives the most intuitive way to define Ising percolation (and inspired
Swendsen and Wang in their proposal~\cite{sw}):
given a generic configuration of the Ising model, delete all the bonds which
join spins with opposite sign and, on the remaining graph, construct a standard
percolation process i.e. switch on the bonds at random with probability $p$.
The resulting graph $G$ will be a typical configuration of an Ising percolation
model.

\section{Cluster structure}

\label{links}
As we have seen, an interesting
 feature of the percolation mapping is that it allows a
geometric characterization of various thermodynamic quantities. In order to
better understand this geometric setting it is convenient to study in a
more precise and detailed way the cluster structure in a typical percolation
configuration.

An important step in this direction was made by Stanley in 1977~\cite{s77}
who noted that at $p=p_c$, in a generic percolating cluster we can
distinguish three different sets of bonds. By associating an electric unit
resistance to each bond, and applying a voltage between the ends of the
cluster, one can select the ``dangling bonds'' (also called
{\sl green} bonds) which are those which do not carry current. The remaining
bonds form the  {\sl backbone}; in this set one can then
select the singly connected bonds ({\sl red} bonds) , which carry the whole
current and have the property that if one is cut then the cluster breaks in
two parts. The remaining bonds are multiply connected and are usually
denoted as {\sl blue} bonds. Starting from this coarse grained
classification, one can then
look at more subtle structures, selecting for instance pairs of double
connected bonds, triples... Equivalently one can look to the subsets of bonds
which carry exactly half of the whole current, one third, and so on.

{\begin{figure}[ht]
\[\begin{array}{ll}
\epsfxsize=.51\linewidth\epsfbox{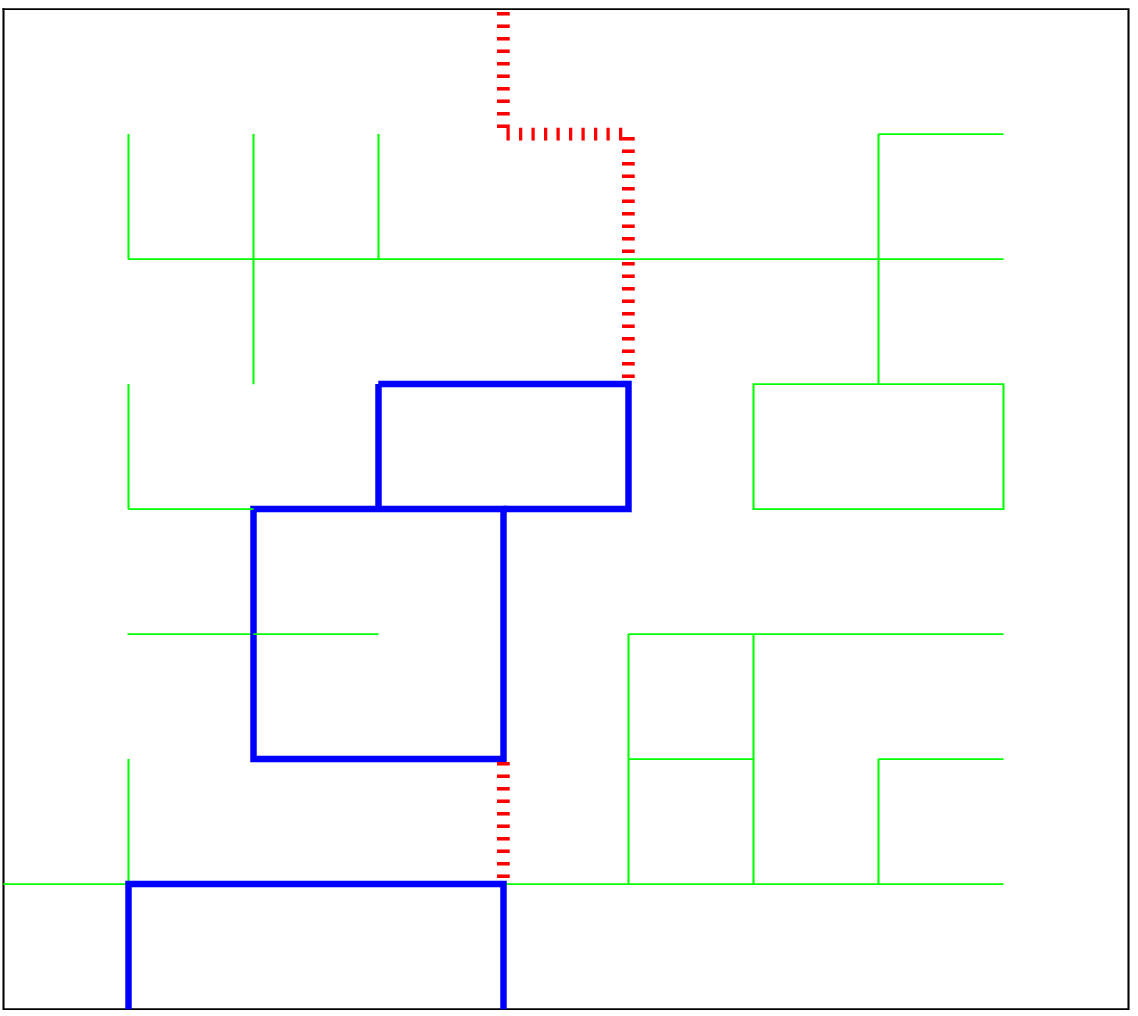}
{}&
\epsfxsize=.51\linewidth\epsfbox{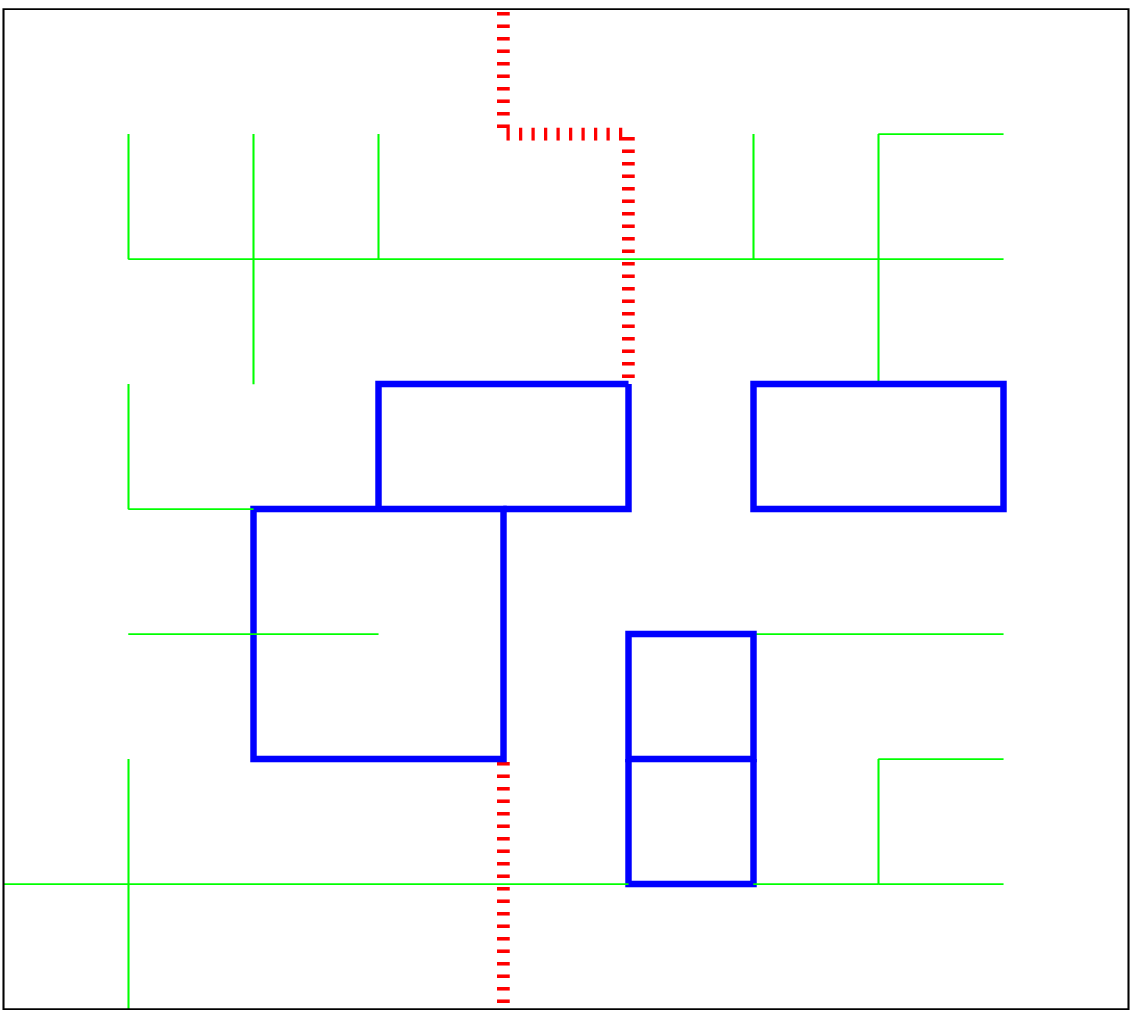}
\\
\parbox[t]{.51\linewidth}{\small 1a) Stanley classification
of the bonds of a percolating cluster. The thin lines denote
the dangling bonds, the thick lines the backbone and the dashed lines  the
red bonds. Free b.c. are understood}
{}~&~
\parbox[t]{.51\linewidth}{\small 1b) The same graph of fig.1a
in the topological classification described in the text.
Thin lines are the grey bonds, thick lines the black bonds and the
dashed lines the red bonds. Periodic b.c. are understood.}
\end{array}\]
\label{cel}
\end{figure}}
\vskip .3 cm
With respect to this standard classification, which assumes free boundary
conditions, in the present work we prefer to
deal with  periodic boundary conditions.
This is because, as we shall see, the new representation of the thermal operators
does not depend explicitly on the percolating property of the configurations,
but rather on their winding and connection properties.
We are then forced to introduce a different partition of the bonds
which  slightly differs from that of Stanley and applies not only to
the (infinite) cluster wrapped along one or more directions, but also
to any FK cluster contributing to the partition function.

We suggest to the reader to follow
our definitions looking at fig.1, where the bonds of a percolating cluster
are split into three sets according to the Stanley classification (fig.1a)
and ours (fig. 2b).

$G$ denotes the generic graph which appears in the sum of eq. (\ref{part4}).
It is composed by  disconnected FK clusters.
We call $G_1$ the set of those bonds which, when deleted,
disconnect the cluster to
which they belong into two disjoint clusters. Remember that we consider
as a cluster also that composed by only one site and no bonds.
We call the bonds of $G_1$  {\sl grey} or {\sl cutting}  bonds.

Let us call $G_0$
the complement of $G_1$ in $G$,
i.e. $G=G_0\cup G_1$.  $G_0$ contains two kinds of
bonds: those which, if deleted,  do not change the wrapping properties of
$G$ are called {\sl black} bonds; those which transform a wrapped graph into
an unwrapped one are called {\sl red} bonds.

It is clear that our grey bonds form a subset of the dangling bonds
of the Stanley's classification, but there are dangling bonds that
are black in our classification, as comparison of fig.s 1a) and 1b)
shows.

Using again an electric circuit analogy, one can say that our black and red
bonds are those which carry current in presence of a variable
magnetic field pointing in a generic direction
\footnote{We thank Lev Shchur for this observation.}.

It is known that the winding properties of the FK clusters are related to
the interface tension in the low-temperature phase \cite{h93}. In this paper
we shall see that the black bonds are directly involved in a new
representation of the internal energy.

Our classification can be
easily iterated by identifying other topological subsets of $G_0$
associated to other thermal observables.
We need in particular to define another set, denoted with $H_2$, that will
play a major role in discussing the new representation of  the specific heat.
$H_2$ is the set of ``cutting pairs'' of $G_0$,  i.e. the set of those
pairs of black bonds which, if simultaneously deleted,
disconnect the cluster to
which they belong into two separate
clusters\footnote{Notice that, since by definition
$G_0$ does not contain cutting bonds, at least two bonds are needed to
disconnect a cluster.}.

Let us see a specific  construction of the set $H_2$, which is interesting both
 because it sheds light on the structure of $H_2$ and because it can be
 straightforwardly iterated to construct ``cutting triplets'' and from them
 possibly further higher combinations which contribute to higher derivatives
 of the free energy (in this paper however we shall not study
these more complicated cases):

\begin{itemize}
\item
Choose a link $l\in G_0$ and
construct the graph $G_0(l)$ obtained from $G_0$ by deleting the
 link $l$.
\item
select in $G_0(l)$ the cutting bonds (as we did above when we
constructed $G_1$ starting from $G$), which form a set $G_1(l)$.
\item
repeat the operation for all $l\in G_0$. The set of pairs $\{(l,h),\forall
h\in G_1(l), \forall l \in G_0 \}$ is exactly twice  $H_2$
(due to the fact that each
pair appears two times).  This means that (apart from a factor two) $H_2$
coincides with the collection of graphs $G_1(l)$ .

\end{itemize}

\section{Ising model with a set of antiferromagnetic links}
\label{frustration}
A natural extension of the Ising model discussed in sect. \ref{ising} is the
one in which we allow the coupling $J$ to change sign from link to link.
Those links for which $J=-1$ are said to be antiferromagnetic or flipped links.
In analogy with the definitions of sect. \ref{ising} (and setting for
simplicity $h=0$) we have now
\eq
 H(J) = - \sum_{\bra n,m\ket }J_{\bra n,m\ket } s_n s_m \; ,
\label{Sspinf}
\en
{}from which we may construct the partition function:
\eq
Z(\beta,J)=\sum_{s_n=\pm1}~e^{-\beta~S(J)}
\label{partf}
\en
We denote the partition function without antiferromagnetic links and with
periodic boundary conditions by $Z(\beta)$. Denoting by $F$ the set
of flipped links we can write
\eq
\frac{Z(\beta,J_F)}{Z(\beta)}= \bra \prod_{\ij\in F}
e^{-2 \beta s_is_j}\ket ~~,
\label{fru}
\en
where the mean values is taken with respect to the standard, unfrustrated
Ising Hamiltonian. It is almost evident that suitable linear
combinations of these expectation values reproduce the whole
even sector of the theory. This can be simply proved using the
Kramers -Wannier duality:
This transformation gives a one-to-one map
between these expectation values and the set of all
correlation functions among the physical observables
of the even sector of dual theory, as described in the next
subsection.

\subsection{Duality and frustrations}
  \label{duality}
The Ising model defined in section \ref{ising} is characterized by the two
following features: the dynamical variables (the spins) live on the $sites$
 and the interaction is defined on the $links$  of the
lattice $\Lambda$. Sites and links are
respectively 0 and 1 dimensional simplexes of the lattice. We can easily
generalize the Ising model by looking to different geometrical realizations of
dynamical variables and interaction Hamiltonians. For instance, in the Ising
gauge model we take the
spins $\tilde{s}_l$ to live on links $l$ and the interaction to be defined
on plaquettes $  {\tilde{s}}_{\bx}=
\prod_{l\in \bx}{\tilde{s}}_l$ .
We could also choose the dynamical variables on the plaquettes and the
Hamiltonian to live on cubes, thus defining a Kalb-Ramond type theory.

It is well known that the Kramers -Wannier duality transformation can be
generalized to Ising models in any dimension $d$. This transformation is not
a symmetry of the model: it maps one description of the dynamical system
to another description of the {\sl same system}. This
duality  transforms a given lattice $\Lambda$ (in our case the (hyper)cubic
lattice on which the Ising model is defined),
into a new lattice (the dual lattice $\tilde\Lambda$) in which each
$k$ dimensional simplex is mapped  to a  $d-k$ one.
In the case of (hyper)cubic lattices the dual lattice turns out to be again
(hyper)cubic.

Under this mapping the Ising model on $\Lambda$ is transformed into a new
model whose Hamiltonian lives on a $d-1$ simplex of $ \tilde\Lambda $.
Thus we have the following correspondences:
\vskip 0.3cm
if $d=2$  \hskip 1cm Ising model $\Longleftrightarrow$ Ising model
\vskip 0.3cm
if $d=3$   \hskip 1cm Ising model $\Longleftrightarrow$  $Z_2$ gauge model
\vskip 0.3cm
if $d=4$   \hskip 1cm Ising model$\Longleftrightarrow$ $Z_2$  Kalb-Ramond
model.
\vskip 0.3cm

\noindent
It can be shown that
in the thermodynamic limit the free energy density of
the original
Ising model coincides (apart from an additive function
of $\beta$ which can be evaluated
exactly) with that of the
dual model  evaluated at the dual coupling $\tilde\beta$
defined as
\eq
\tilde\beta=-\frac12 ~\log[th(\beta)]~~~.
\label{dualbeta}
\en
Using a duality transformation it is possible to build up a one-to-one
mapping between a given pattern $J=\{J_{\ij}\}$ of antiferromagnetic links
and the correlators of the physical observables of the dual description.
Take for instance the 3$D$ case and denote by $F$ the set of
flipped links. Associate to each of these links the corresponding
plaquette of the dual lattice and denote by $\tilde{F}$ this set.
Duality implies that
\footnote{It is important to stress that eq.(\ref{frudu}) holds only in the
thermodynamic limit, in which the boundary conditions can be neglected.
In fact in general the duality transformation does not preserve
 specific choices of the boundary conditions. In particular, periodic b.c.
are mapped by duality into a mixture of periodic and antiperiodic
boundary conditions.}

\eq
\bra \prod_{\ij\in F}e^{-2\beta s_is_j}\ket =\bra \prod_{\bx\in\tilde{F}}
\tilde{s}_{\bx}\ket _{\rm gauge}~~,
\label{frudu}
\en
where the left-hand side coincides with the ratio of the partition
functions defined in eq.(\ref{fru}), while the right-hand side
is the expectation value of the product of plaquette variables
$\tilde{s}_{\bx}$ with a Boltzmann factor $e^{-\tilde\beta S_{\rm gauge}}$
with
\eq
S_{\rm gauge}=-\sum_\bx \tilde{s}_\bx~,
~~~\tilde{s}_\bx=\tilde s_{l_1}\tilde s_{l_2}\tilde s_{l_3}\tilde s_{l_4},
\en
where $\tilde s_l\in\{1,-1\}$ are the variables located in the links
of $\tilde\Lambda$.

Conversely, given any product of Wilson loops $W(C1)W(C2)\dots$,
we can replace it (in many ways) with an equivalent set of elementary
plaquettes, with the only constraint that the boundary of this set
should coincide with the set of loops $\{C1,C2,\dots\}$, then we can
apply again eq.(\ref{frudu}) and this completes the proof of the
above-mentioned one-to-one correspondence in the 3$D$ case. The
extension to other dimensions is straightforward.

We want now to use this correspondence to show that for the even sector of the
spin Ising model (namely for the operators obtained as  products of an even
number of spins and hence invariant under the $Z_2$ symmetry of the model)
 the set of ratios
(\ref{fru}) corresponding to all the possible choices of antiferromagnetic
bonds forms a
complete set in the sense that any correlator of the spin Ising model
can be expressed in terms of these ratios. To be definite we put
forward this argument again for the 3$D$ case, its generalization
to other dimensions being straightforward. The set of  products of
Wilson loops forms a complete set of observables of the dual description;
the simplest, non-trivial observable is the plaquette. The spin-spin
correlators on the original lattice $\Lambda$ can be expressed in terms of the
gauge model by flipping  the set of plaquettes of
$\tilde\Lambda$  intersected by a path $\gamma\subset\Lambda$ connecting the
site $x$ to the site $y$:
\eq
\bra s_xs_y\ket =\bra \prod_{\bx\in\gamma}
e^{-2\tilde\beta\tilde{s}_\bx}\ket _{\rm gauge}
\en

Using the obvious identity
$ e^{-2\tilde\beta\tilde s_\bx}=\cosh 2\tilde\beta-\sinh 2\tilde\beta
\tilde s_\bx $
the right-hand side can be written as a combination of expectation
values of products of Wilson loops. Then using eq.(\ref{frudu}) we
can evaluate this correlator directly in the Ising model in terms
of flipped partition functions. Such a construction can be repeated
for any other spin correlator and easily generalized to any
dimension $d$.  In conclusion, we can state that the whole set of spin
correlators of the Ising model in any dimension can be encoded in a
suitable set of ratios of flipped partition functions.

\section{Percolation mapping of the frustrated model}
\label{fruma}
The percolation mapping discussed in sect. \ref{map} can be extended also
to the
frustrated Ising model~\cite{clmp}~\footnote{See
also~\cite{sg} for some recent application of the percolation mapping in
presence of frustrations to the study of disordered systems.}. The main
difference with respect to the unfrustrated case is that now
also the topology of the cluster is important. Let us see this
mapping in detail.

 We assume that the set of couplings $J_{ij}$ of
 eq.(\ref{Sspin}) is given by an arbitrary, fixed
 collection of signs $ \pm 1$.

As in the  case without frustrations the Ising partition function of
eq.(\ref{partf}) can be rewritten as:
\eq
Z(\beta,J)=e^{N\beta}\sum_{s_i=\pm1}\prod_{\ij}
\left[1-p+p\,\delta(J_{\ij}s_is_j)\right]
\label{part2bis}
\en
 where
 $\delta(J_{\ij}s_is_j)\equiv(1+J_{\ij}s_is_j)/2$ is a projector on
 configurations
 with $J_{\ij}s_is_j=1~$.
Expanding the products in eq.(\ref{part2bis}) we find:
\eq
Z(\beta,J)=e^{N\beta}\sum_{G}(1-p)^{N-N(G)}p^{N(G)}
\sum_{s_i=\pm1}\left[\prod_{\ij\in G} \delta(J_{\ij}s_is_j)\right]
\label{part3bis}
\en
as in sect. \ref{ising}, $N$ is the total number of links of the lattice,
 the summation goes over all the subgraphs  $G$ of the lattice,
$N(G)$ is the number of links of  $G$ and $p=1-e^{-2\beta}$.

Note that the whole dependence on the signs $J_{\ij}$ is contained in
the $ \delta$ projectors and that the product inside the square
brackets is a projector which forces all the links of each connected
component $G_c$ of the graph to fulfill the constraint
\eq
J_{\ij}s_is_j=1~~~,~~~\ij\in G_c~~.
\label{constr}
\en
We say that a cluster $G_c$ is {\sl compatible} with a given choice of
couplings $J_{\ij}$ if there is a configuration of its sites obeying
such a constraint. In the standard case $(J_{\ij}=1~ \forall ~ \ij)$ this
condition implies simply that the sites of each cluster have the same sign.
In presence of frustrations it is easy to verify that eq.(\ref{constr})
is a topological constraint, telling us that a connected graph $G_c$ is compatible
if and only if no loop of $G_c$ includes an odd number of antiferromagnetic links.

A crucial observation is that, owing to the
connected nature of the cluster, the value of $s_i$ of any site $i$ of a
compatible cluster fixes the values of all the other sites of the cluster,
and that if $ \{ s_i~,~i\in G_c \} $ is a solution of the constraint,
also the opposite configuration $\{-s_i \}$  is a solution. Thus any
connected subgraph  contributes to the partition function with a
factor of 2 if it is a compatible cluster, otherwise it gives a zero
contribution.
Thus summing on the spin configurations in eq.(\ref{part3bis}) we end up with

\eq
Z(\beta,J)=e^{N\beta}\sum_{G}~ \varpi_J(G)
p^{N(G)}(1-p)^{N-N(G)}~2^{C(G)}~~~.
\label{part4bis}
\en
where  $ \varpi_J$ denotes the projector on
compatible graphs, $i.e.$ graphs made with compatible clusters,
defined as follows

\eq
\varpi_J(G)=\left\{\begin{array}{ll}$1$ & \mbox{if no loop of $G$ contains an odd}\\
   $~$& \mbox{ number of antiferromagnetic links} \\
$0$ & \mbox{otherwise~.}\end{array}\right.
\label{improved}
\en

\vskip .3 cm
In the standard, unfrustrated case the sum over $G$ is unconstrained and we
obtain the result discussed in sect. \ref{map}.
When there are  frustrations, the set of compatible graphs is a
proper subset of the all the possible subgraphs of the lattice. We
can then write the following exact relation, which is the cluster
version of the eq.(\ref{fru}):
\eq
\frac{Z(\beta,J)}{ Z (\beta)}=\bra \varpi_J\ket ~~,
\label{main}
\en
where the expectation value is taken with respect the standard
Hamiltonian.

An interesting, particular example is the  Ising model on a cubic lattice
with periodic $(p)$ boundary conditions in one coordinate direction,
say $z$, in which all the links of a slice orthogonal to $z$ are flipped
to  $-1$. Such a pattern of antiferromagnetic couplings is equivalent to
choose antiperiodic b.c. $(a) $ along the $z$ direction. Denoting with
$Z_a$ and $Z_p$ the partition functions with these two different choices
of b.c., we have
\eq
\frac{Z_a}{Z_p}=\bra\varpi_z\ket~~,
\label{has}
\en
where $\varpi_z$ is a projector on the FK graphs which is 1 if there
is no cluster with an odd winding number in the $z$ direction and 0
otherwise. Eq.(\ref{has}) is the starting point of a new representation
of the interface free energy first found in ref. \cite{h93} which is
particularly useful near the critical point.

Coming back to the general formula (\ref{main}), we would like to make
a few comments:

\begin{itemize}
\item Eq.(\ref{main}) is valid not only for any regular lattice in any
dimension, but also for the Ising model defined on an arbitrary graph
and for any choice of antiferromagnetic couplings.

\item Any correlator of the even sector of the Ising model can be
expressed in terms of the ratios (\ref{main}).
\item
The projector $\varpi_J$ depends only on the subgraph $G_0\subset G$.
\end{itemize}
As a consequence,
there is no information loss in the even sector if one delete all the grey
bonds of any graph $G$, provided  one use eq.(\ref{main}) to evaluate
these observables.
In other words, in the partition function (\ref{part4}) we can split the
sum over all possible graphs $\sum_G$ as the double sum
\eq
Z(\beta)=e^{N\beta}\sum_{G_0}\sum_{G_1(G_0)}~p^{N(G)}(1-p)^{N-N(G)}~2^{C(G)}
\equiv\sum_{G_0}e^{-{\cal H}(G_0;\beta)}~,
 \label{blue}
 \en
where $G=G_0\cup G_1$ and $G_1(G_0)$ denotes an arbitrary
set of grey bonds compatible with a fixed set $G_0$ of
black (and red) bonds. $\cal H$ defines a new Hamiltonian
which depend only on the configurations of the black bonds.
Eq. (\ref{main}) tell us that, in spite of the
 sum over all the possible insertions of grey bonds,
in the resulting Hamiltonian ${\cal H}$ it is encoded exactly the same
piece of information about the even sector than that of the original Ising
Hamiltonian. Though this fact will not be exploited in the present
paper, it clearly suggests the existence of an hidden huge symmetry
of the theory.

\section{Cluster description of thermal observables}
\label{thermal}
We are at this point in the position to study a new cluster representation of
thermal observables, alternative to the one presented at the end of
sect. \ref{map}.
Let us look first at the internal energy.

\subsection{Internal energy}
\label{energy}
Let us consider an Ising system defined  on an arbitrary lattice in
$d$ space dimensions
 with only one antiferromagnetic link in the
position $\ij$. Then eq.(\ref{fru}) can be rewritten explicitly as
\eq
\frac{Z(\beta,J_{\ij})}{Z(\beta)}= \bra
e^{-2 \beta s_is_j}\ket\,=\,\cosh 2\beta -\bra s_is_j\ket \sinh 2\beta
~~.
\label{one}
\en
On the other hand, using eq.(\ref{main}) we get
\eq
\bra s_is_j\ket={\rm cotanh\,}2\beta- \frac{\bra \varpi_{\ij}\ket}
{\sinh 2\beta}~~,
\label{two}
\en
where $\varpi_{\ij}$ is the projector on the graphs compatible with
the antiferromagnetic link in the position $\ij$.
Such graphs  are of two types:
\begin{itemize}
\item
those in which the link $\bra ij\ket $ does not appear;
\item
those  in which the cluster which contains the bond $\ij $
is split into two separate clusters when it  is deleted.
\end{itemize}
These two conditions tell us simply that $\ij $ does not belong to the
subgraph $G_0$ of $G$.
 Summing over all the possible links and taking advantage of the
 translational invariance of the lattice we find:
\eq
\bra \sum_{\bra n,m\ket }\varpi_{\bra nm\ket }\ket =N-\bra N(G_0)\ket ~~~.
\en
Combining  this result with eq.s(\ref{two}) and (\ref{intene}) we get
the sought-after representation of the energy in terms of black bonds
\eq
E=\frac{ \bra N(G_0)\ket }{\sinh(2\beta)}+N\,\tanh \beta~,
\label{aa4}
\en
as anticipated in the introduction. This is the first main outcome
of our analysis.
%Comparison with eq.s (\ref{ie1}) and (\ref{bb1}) yields
%\eq
%y(G_0)=y(G)=d~~~.
%\en

\subsection{Specific heat}
  \label{heat}
Let us start from the following definition of the specific heat

\begin{equation}
C =
\bra (\sum_{\bra n,m\ket } s_n s_m)(\sum_{\bra k,l\ket } s_k s_l) \ket ~-~\bra \sum_{\bra n,m\ket } s_n s_m)\ket ^2
\label{calspecbis}
\end{equation}

call $f=\bra nm\ket$ and $g=\bra kl\ket$ the two links and
separate the case $f=g$ in the sum. We obtain

\begin{equation}
C = (N-\frac{E^2}{N})+ \sum_{f\not=g}\left[
\bra s_fs_g\ket ~-~\bra s_f\ket \bra s_g\ket \right]~,
\label{dd1}
\end{equation}
with $s_f\equiv s_ms_n$ and $s_g\equiv s_ks_l$.

In order to get the new cluster representation of $C$ it is sufficient
to consider now a system with two antiferromagnetic links located in
$f$ and $g$ with the associated projector $\varpi_{fg}$. Then we have
\eq
\bra\varpi_{fg}\ket\equiv
\bra e^{-2\beta (s_f+s_g)}\ket=\bra(\cosh 2\beta -s_f\sinh2\beta)
(\cosh 2\beta-s_g\sinh2\beta)\ket~.
\label{twolink}
\en
Using again eq.(\ref{two}) we get
\begin{equation}
 \sum_{f\not=g}\left[
\bra s_fs_g \ket ~-~\bra s_f\ket \bra s_g\ket \right]=
 \sum_{f\not=g}
\frac{\bra \varpi_{fg}\ket -
\bra \varpi_{f}\ket \bra \varpi_{g}\ket }{\sinh^22\beta}~.
\label{dd2}
\end{equation}

The graphs with $\varpi_{fg}=1$ are of two types:
\begin{itemize}
\item
those in which both  $f$ and $g$  do not belong to the
$G_0$ subgraph; when summing over $f$ and  $g$ these configurations
give simply the contribution
$\sum_{fg}\varpi_f\varpi_g=(N-N(G_0))(N-N(G_0)-1)$.
\item
those  in which both $f$ and $g$ belong to $G_0$ but are such that
any loop going through $f$ contains also $g$.
Let us denote by ${\cal G}_{fg}$ the projector which selects the
 graphs  with this
special property.
\end{itemize}
We may give a cluster representation of the specific heat if we are able to
evaluate the sum $\sum_{f\not=g}{\cal G}_{fg}$.
 Reversing the order of
 summations, i.e. taking first the sum over all the pair of links and then
 the sum over  $G$ implied in the expectation value we obtain
 \eq
 \sum_{f\not=g}\bra {\cal G}_{fg}\ket=2\bra N(H_2)\ket
 \label{ting}
 \en
In fact, from the definition of  ${\cal G}_{fg}$, if
 ${\cal G}_{fg}=1$ then cutting simultaneously
 $f$ and $g$ we disconnect the cluster to which they belong into two
 subclusters, and conversely if two links belong to $G_0$ and are a cutting pair
then they certainly fulfill the condition ${\cal G}_{fg}=1$. The factor two is
due to the fact that in the sum the pair $f,g$ appears twice.

Collecting together the various pieces we end up with
\eq
C = (N-\frac{E^2}{N})+ \frac{2 \bra N(H_2)\ket+\sum_{f\not=g}(\bra\varpi_f\varpi_g\ket
-\bra\varpi_f\ket\bra\varpi_g\ket)}{\sinh^2(2\beta)}
\en
which gives the new representation of the specific heat in terms of
black bonds.

\section{Conclusions}
\label{conclu}
In this paper we have shown how to construct a new representation of thermal
operators in Ising percolation in terms of a new set of bonds, called the
black bonds, forming a subset
of the standard FK clusters. Our approach is quite general, indeed we never
needed to specify the lattice structure. Actually our results are
true also for Ising models defined on arbitrary graphs.

Our main results are:
\begin{itemize}
\item
We  proposed a new scheme to classify the bonds
of a cluster on the basis of their topological properties.
 We have discussed the relations with the standard classification
scheme.

\item
We pointed out  that in the
 Ising model all the correlators  of an even
number of spins can be expressed as ratios of partition functions with
flipped links. In particular we have applied this result to the
internal energy and the specific heat.
\item
We  found that in the configurations of black bonds contributing
to the partition function is encoded the whole
information on the even sector of the theory; in particular the
internal energy can be expressed in terms of the
mean number of black bonds and that the specific heat is related
to their variance and to the special subset $H_2$ defined in sect.
\ref{links}.

\item
finally we observed that our new representation of observables of the
even sector suggests the existence of an hidden huge symmetry of the
Ising model.
\end{itemize}

Our results can be straightforwardly extended to $q$-state Potts models with
generic values of $q$ \cite{cgn}.

The most relevant application of the present analysis is that it may help to
find new powerful algorithms to simulate the Ising model or to construct
improved estimators for thermal observables. As a matter of fact for some
special configurations of antiferromagnetic links these algorithms already exist.
 In the case of a whole flipped hyperplane  leading to antiperiodic
 boundary conditions this possibility was discussed for the first
 time by  M.Hasenbusch in~\cite{h93}, who provided a very
  powerful tool to evaluate the surface tension. Later,
 it was modified so as to evaluate in the dual version
 Wilson loops \cite{cfghp}, correlators of Polyakov loops~\cite{gv} and
  plaquette expectation values~\cite{gp} in the gauge Ising
  model. We hope that the present analysis could help to further extend the
  range of these applications.

\vskip 1cm

{\bf  Acknowledgements}
 We warmly thank A.Coniglio, M. Hasenbusch, L.N. Shchur and A.Sokal for several
useful suggestions, which helped us to write the present version of this
paper. In particular we are deeply indebted to A.Sokal
for pointing us an error in  our previous derivation of the
cluster representation for the specific heat.

This work is supported in part by the European Commission TMR programme
ERBFMRX-CT96-0045.

\end{document}